\documentclass[twocolumn,showpacs,prb,amsmath,amssymb,superscriptaddress]{revtex4-1}

\usepackage{epstopdf}
\usepackage{graphicx}% Include figure files
\usepackage{dcolumn}% Align table columns on decimal point
\usepackage{bm}% bold math

\begin{document}

\preprint{APS/123-QED}

\title{Role of the Ce valence in the coexistence of superconductivity and ferromagnetism of CeO$_{\bm{1-x}}$F$_{\bm{x}}$BiS$_{\bm{2}}$ 
revealed by Ce $\bm{L_3}$-edge x-ray absorption spectroscopy}

\author{T. Sugimoto}
\affiliation{Department of Complexity Science and Engineering, University of Tokyo, 5-1-5 Kashiwanoha, Chiba 277-8561, Japan}

\author{B. Joseph}
\affiliation{Dipartimento di Fisica, Universit\'a di Roma ``La Sapienza'' - Piazzale Aldo Moro 2, 00185 Roma, Italy}

\author{E. Paris}
\affiliation{Dipartimento di Fisica, Universit\'a di Roma ``La Sapienza'' - Piazzale Aldo Moro 2, 00185 Roma, Italy}

\author{A. Iadecola}
\affiliation{Elettra - Sincrotrone SCpAss 14 - km 163.5 34149 Basovizza, Trieste, Italy}

\author{T. Mizokawa}
\affiliation{Department of Complexity Science and Engineering, University of Tokyo, 5-1-5 Kashiwanoha, Chiba 277-8561, Japan}
\affiliation{Dipartimento di Fisica, Universit\'a di Roma ``La Sapienza'' - Piazzale Aldo Moro 2, 00185 Roma, Italy}
\affiliation{Department of Physics, University of Tokyo, 5-1-5 Kashiwanoha, Chiba 277-8561, Japan}

\author{S. Demura}
\affiliation{National Institute for Materials Science, 1-2-1 Sengen, Tsukuba 305-0047, Japan}
%\affiliation{JST-EU-JAPAN, 1-2-1 Sengen, Tsukuba, 305-0047, Ibaraki, Japan}
\affiliation{Graduate School of Pure and Applied Science, University of Tsukuba, 1-1-1 Tennodai, Tsukuba 305-8577,  Japan}

\author{Y. Mizuguchi}
\affiliation{Department of Electrical and Electronic Engineering, Tokyo Metropolitan University, 1-1 Minami-osawa, Hachioji 192-0397, Japan}

\author{Y. Takano}
\affiliation{National Institute for Materials Science, 1-2-1 Sengen, Tsukuba 305-0047,  Japan}
%\affiliation{JST-EU-JAPAN, 1-2-1 Sengen, Tsukuba, 305-0047, Ibaraki, Japan}
\affiliation{Graduate School of Pure and Applied Science, University of Tsukuba, 1-1-1 Tennodai, Tsukuba 305-8577,  Japan}

\author{N. L. Saini}
\affiliation{Dipartimento di Fisica, Universit\'a di Roma ``La Sapienza'' - Piazzale Aldo Moro 2, 00185 Roma, Italy}

\date{\today}% It is always \today, today,

\begin{abstract}
We have performed Ce $L_3$-edge x-ray absorption spectroscopy (XAS) measurements on CeO$_{1-x}$F$_x$BiS$_2$, in which the superconductivity of the BiS$_2$ layer and the ferromagnetism of the CeO$_{1-x}$F$_x$ layer are induced by the F-doping, in order to investigate the impact of the F-doping on the local electronic and lattice structures. The Ce $L_3$-edge XAS spectrum of CeOBiS$_2$ exhibits coexistence of $4f^1$ (Ce$^{3+}$) and $4f^0$ (Ce$^{4+}$) state transitions revealing Ce mixed valency in this system. The spectral weight of the $4f^0$ state decreases with the F-doping and completely disappears for $x>0.4$ where the system shows the superconductivity and the ferromagnetism. The results suggest that suppression of Ce-S-Bi coupling channel by the F-doping appears to drive the system from the valence fluctuation regime to the Kondo-like regime, leading to the coexistence of the superconducting BiS$_2$ layer and the ferromagnetic CeO$_{1-x}$F$_x$ layer.
\end{abstract}

\pacs{74.25.Jb, 74.70.Xa, 78.70.Dm, 71.28.+d}% PACS, the Physics and Astronomy
\maketitle

\newpage

%\section{Introduction}

Since the discovery of the high-$T_c$ superconductivity in 
the layered Fe pnictides, \cite{0} tremendous research efforts
have been dedicated to explore new superconductors in various pnictides and chalcogenides with layered structures.
Very recently, a new family of superconductors with the BiS$_2$ layer
has been discovered by Mizuguchi {\it et al.}, \cite{1} and 
the discovery followed by the extensive research activities on the related systems including REO$_{1-x}$F$_x$BiS$_2$ (RE = rare-earth elements) with the BiS$_2$ layer.
\cite{2,3,4,5,6,7,8,9,10,11,12,13,14,15,16}
In the REO$_{1-x}$F$_x$BiS$_2$, the electronically active BiS$_2$ layers are separated by the REO spacers, and the band filling and the superconductivity of the BiS$_2$ layer can be controlled by the F-doping in the REO spacers. The maximum $T_c$ of 10.5 K has been achieved in LaO$_{1-x}$F$_{x}$BiS$_2$ synthesized by high pressure annealing \cite{5} whereas that synthesized at ambient pressure tends to be non-superconducting.
In addition, high pressure studies on optimally doped LaO$_{0.5}$F$_{0.5}$BiS$_2$ (synthesized at ambient pressure) have revealed that $T_c$ increases up to the maximum value of 10.1 K at 1-2 GPa  followed by a gradual decrease with increasing pressure.
\cite{17,18,19} These experiments indicate that the superconductivity
in BiS$_2$-based superconductors is highly sensitive to the local atomic displacements.

While various BiS$_2$-based systems have been synthesized with different spacer layers, the particular case of CeO$_{1-x}$F$_x$BiS$_2$ with coexistence of superconductivity and ferromagnetism at low temperature is interesting.\cite{20} In fact, CeO$_{1-x}$F$_x$BiS$_2$ shows superconductivity and ferromagnetism with a maximum $T_c$ of 6 K. While BiS$_2$-based superconductors have been considered as conventional superconductors with electron-phonon coupling, the coexistence of superconductivity and ferromagnetism provokes further studies to understand interaction of different electronic degrees of freedom and the role of spacer layers in these materials. In addition, the ferromagnetism and the superconductivity are enhanced in the samples synthesized by high pressure annealing, indicating high sensitivity to the local atomic displacements.

Ce $L_3$ x-ray absorption spectroscopy (XAS) is a direct probe of the local structure around a selected absorbing atom and distribution of the valence electrons, with the final states in the continuum being due to multiple scattering resonances of the photoelectron in a finite cluster. \cite{20a} In this work, we have exploited Ce $L_3$ XAS to investigate the impact of the F-doping  on the local electronic and lattice structures of CeO$_{1-x}$F$_x$BiS$_2$ system synthesized differently, i.e. as grown (AG) and high pressure (HP) annealed. Since it is difficult to prepare clean surfaces using the available polycrystalline CeO$_{1-x}$F$_x$BiS$_2$ samples, the bulk-sensitive Ce $L_3$-edge XAS taken in the transmission mode is the most reliable tool to evaluate the Ce valence.
The Ce $L_3$-edge XAS spectrum of CeOBiS$_2$ exhibits the transitions to the $4f^1$ and $4f^0$ final states, indicating the Ce$^{3+}$/Ce$^{4+}$ valence fluctuation that should be related with the Ce-S-Bi coupling channel. Namely, the Ce$^{3+}$ ($4f^1$) and Ce$^{4+}$ ($4f^0$) states are mixed in the ground state due to the hybridization between the Ce 4$f$ orbitals and the Bi 6$p$ conduction band.
The spectral weight of the $4f^0$ states decreases with the F-doping and disappears completely for $x> 0.4$ where the system shows the coexistence of the superconductivity and ferromagnetism at low temperature. The F-doping is expected to break the Ce-S-Bi coupling channel and induce crossover from the
valence fluctuation regime to the Kondo-like regime. 
%The localization of the Ce 4$f$ electrons is consistent with the coexistence of the superconductivity in the BiS$_2$ layer and the ferromagnetism in the CeO$_{1-x}$F$_x$ layer.

%\section{Method}
Ce $L_3$-edge XAS measurements were performed on polycrystalline samples of CeO$_{1-x}$F$_x$BiS$_2$ prepared by the solid-state reaction method. Both AG and HP annealed samples were used for the measurements. All the samples were well characterized for their transport and average structural properties prior to the absorption measurements. Details on the sample preparation and characterization are given elsewhere. \cite{20} The x-ray absorption experiments were performed at the XAFS beamline of the Elettra Synchrotron Radiation Facility, Trieste, where the synchrotron radiation emitted by a bending magnet source was monochromatized using a double crystal Si(111) monochromator. The measurements were taken at room temperature in transmission mode using three ionization chambers mounted in series for simultaneous measurements on the sample and a reference. As a routine experimental approach, several absorption scans were collected on each sample to ensure the reproducibility of the spectra, in addition to the high signal to noise ratio.

%\section{Results and Discussion}

\begin{figure}
\includegraphics[width=8cm]{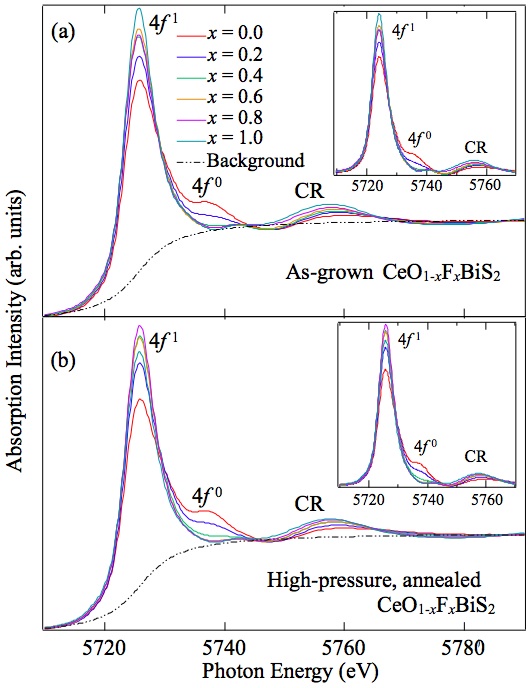}
\caption{(Color online) (a) Ce $L_3$-edge normalized XAS spectra of AG and (b) HP CeO$_{1-x}$F$_x$BiS$_2$ 
with $x=0.0$, 0.2, 0.4, 0.6, 0.8, and 1.0. The $4f^1$, $4f^0$, and CR peaks 
around 5725 eV, 5737 eV, and 5758 eV are shown. The background (arctangent function)-subtracted spectra are shown in the insets. The spectra were taken at room temperature with transmission mode.
}
\label{CeLEdge}
\end{figure}

\begin{figure}
\includegraphics[width=8cm]{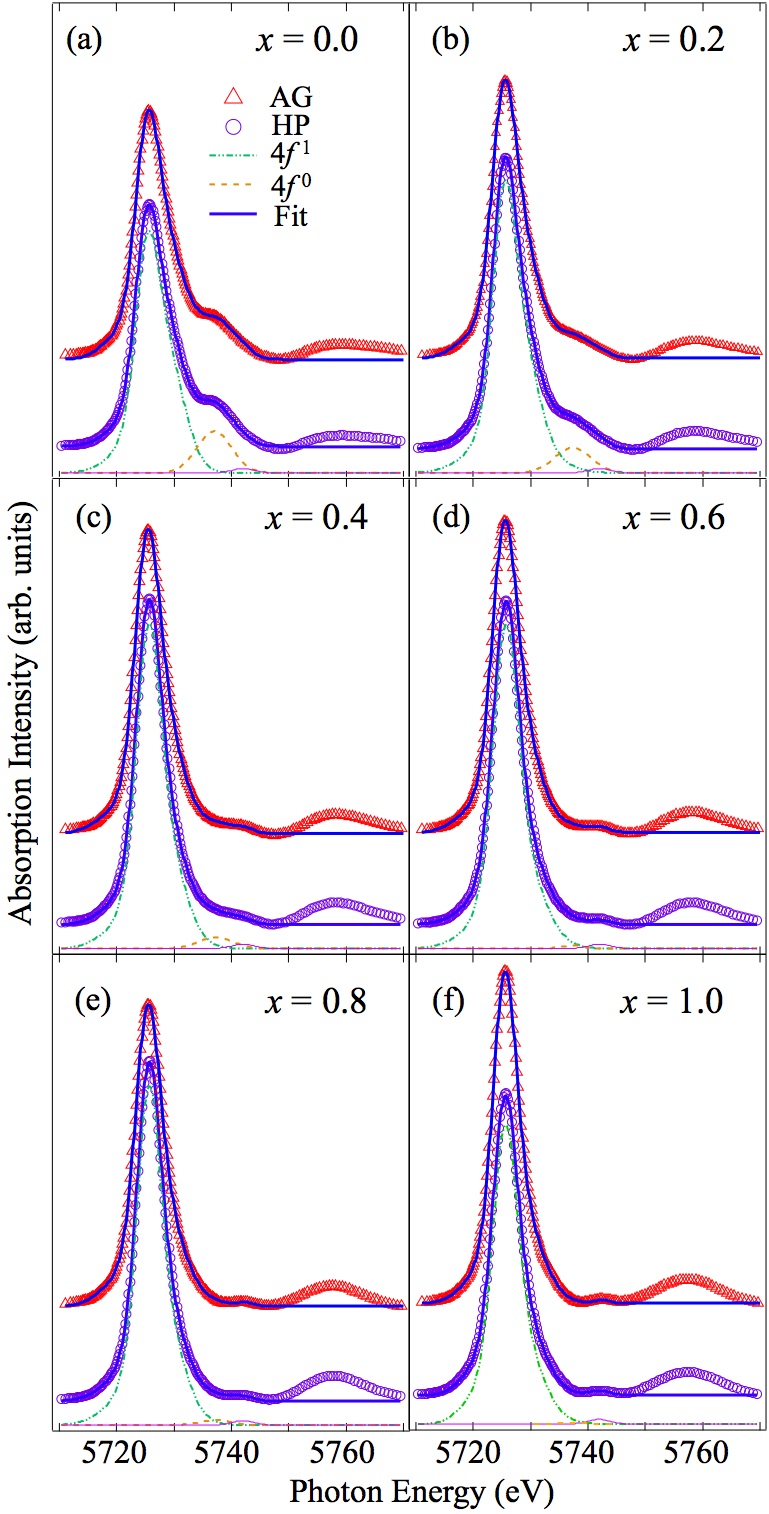}
\caption{(Color online) Multi-curve fitting results on Ce $L_3$-edge XAS spectra in CeO$_{1-x}$F$_x$BiS$_2$ with 
(a) $x=0.0$, (b) $x=0.2$, (c) $x=0.4$, (d) $x=0.6$, (e) $x=0.8$, and (f) $x=1.0$. 
The experimental data of AG and HP are shown as triangles and circles, respectively.  
The fitted results are shown by solid lines on the HP and AG experimental data. Each component of the fitting lines is shown only for the HP experimental data.
The tiny peak around 5742 eV is fixed for all the cases from (a) to (f).}
\label{MultiCurveFit}
\end{figure}

Figure \ref{CeLEdge} displays the Ce $L_3$-edge XAS spectra of AG (Fig. \ref{CeLEdge} (a)) and HP (Fig. \ref{CeLEdge} (b)) CeO$_{1-x}$F$_x$BiS$_2$ ( $x=0.0$, 0.2, 0.4, 0.6, 0.8, and 1.0).
The spectra are normalized with respect to the atomic absorption estimated by a linear fit to the high energy part of the spectra. Three main structures around 5725 eV, 5737 eV, and 5758 eV can be identified on the Ce $L_3$-edge XAS spectra.  The first peak around 5725 eV is the absorption white line corresponding to the transition from the Ce 2$p$ core level to the vacant Ce 5$d$ state mixed with the Ce $4f^1$ final state. \cite{21,22,23,24a} 
%which is stabilized by the attractive Coulomb interaction between the Ce 2$p$ hole and the Ce 4$f$ electron.
On the other hand, the second peak around 5737 eV corresponds to the transition from
the Ce 2$p$ core level to the vacant Ce 5$d$ state mixed with the Ce $4f^0$ final state.
The $4f^1$ and $4f^0$ final states are the so-called well-screened and poorly-screened states
and provide information on the Ce valence states. Presence of both $4f^1$ and $4f^0$ states suggest the Ce$^{3+}$/Ce$^{4+}$ valence fluctuation.
On the other hand, the energy difference between the $4f^1$ and $4f^0$ absorption peaks, 
which is approximately 12 eV \cite{21,22,23} is mainly determined by the Ce 2$p$ - Ce 4$f$ Coulomb interaction and is expected to be independent of the F-doping. One can see a systematic change due to the F-doping in the $4f^1$ and $4f^0$ peak intensity. 
The third peak around 5758 eV includes the information on the local lattice structures. This peak is so-called continuum resonance (CR), likely to be due to Ce-Bi scattering with a contribution from the Ce-Ce scattering, reflecting evolution of the Ce-Bi/Ce bond length. In addition, there is a weak feature around 5742 eV. This feature is  a characteristic feature of layered rare-earth systems, and its intensity is generally sensitive to the O/F atom order/disorder in the CeO/F layer. 

\begin{figure}
\includegraphics[width=8cm]{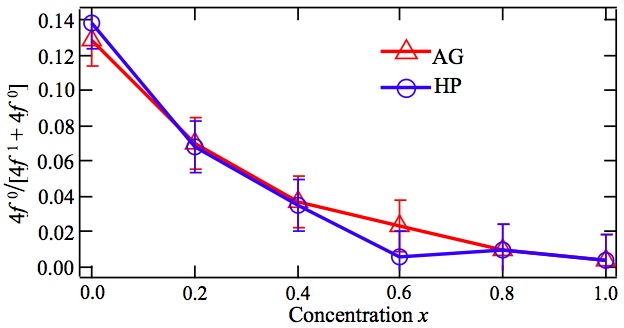}
\caption{(Color online)  
The relative spectral weight of $4f^0$ states for both AG and HP samples. }
\label{ratio}
\end{figure}

In order to qualify the electronic states, we have estimated the spectral weight of $4f^1$ and $4f^0$ absorption peaks using the following procedure. A constant background estimated by an arctangent function was subtracted from the XAS spectra (inset of Fig. \ref{CeLEdge}). The background-subtracted spectra were fitted by Gaussian functions, reproducing all the main peaks, i.e., $4f^1+4f^0$ as shown in Fig. \ref{MultiCurveFit}. 
As for the $4f^1$ peak, we utilized three Gaussian functions to reproduce the
asymmetric line shape, while a symmetric Gaussian function was enough to describe the $4f^0$ peak.  The $4f^0$ peak is found to decrease monotonically with the F-doping. The fit of the weak peak (peak around 5742 eV, characteristic feature of layered structures) is fixed for all the fits since this feature is independent of the F-doping. 
The weak peak was used in the fit in order to ensure the proper area estimation of the $4f^0$ peak.
The CR was not considered in the fit. The fit for all the samples are shown along with the experimental results.

The intensity of the main peaks $4f^1+4f^0$ depends on the amount of additional background from the grain boundaries and the hybridization between the Ce 4$f$ and Ce 5$d$ orbitals which can be changed by the F-doping and the high pressure synthesis.
Therefore, we employ the relative spectral weight $4f^0/[4f^1+4f^0]$, which is obtained by integrating the Gauss functions and is shown in Fig. \ref{ratio}, 
in order to discuss the Ce valence. 
The relatively large $4f^0/[4f^1+4f^0]$ value in CeOBiS$_2$ indicates that 
the Ce$^{4+}$ state with $4f^0$ electronic configuration
is coexisting with the Ce$^{3+}$ state with $4f^1$ electronic configuration. \cite{20,21,22}
The existence of the $4f^0$ peak, namely, the valence fluctuation between 
the Ce$^{3+}$ and Ce$^{4+}$ states in CeOBiS$_2$ is in sharp contrast to
the pure Ce$^{3+}$ state in iron-based CeOFeAs pnictides. \cite{24a,24b} 
The $4f^0/[4f^1+4f^0]$ value decreases with the F-doping both in the AG and HP
samples. The $4f^0/[4f^1+4f^0]$ value of the AG sample is slightly larger than 
the HP sample at $x$=0.6, which is located between the Ce$^{3+}$/Ce$^{4+}$ valence
fluctuation regime and the Ce$^{3+}$ Kondo-like regime. 
The small $4f^0/[4f^1+4f^0]$ value at $x$=0.6 in the HP sample would be 
consistent with the fact that the superconductivity and ferromagnetism 
tend to be enhanced in the HP samples. However, the difference in $4f^0/[4f^1+4f^0]$
is very subtle, suggesting that some additional factors such as inhomogeneity play important role in the difference between the AG and HP samples.

\begin{figure}
\includegraphics[width=7.5cm]{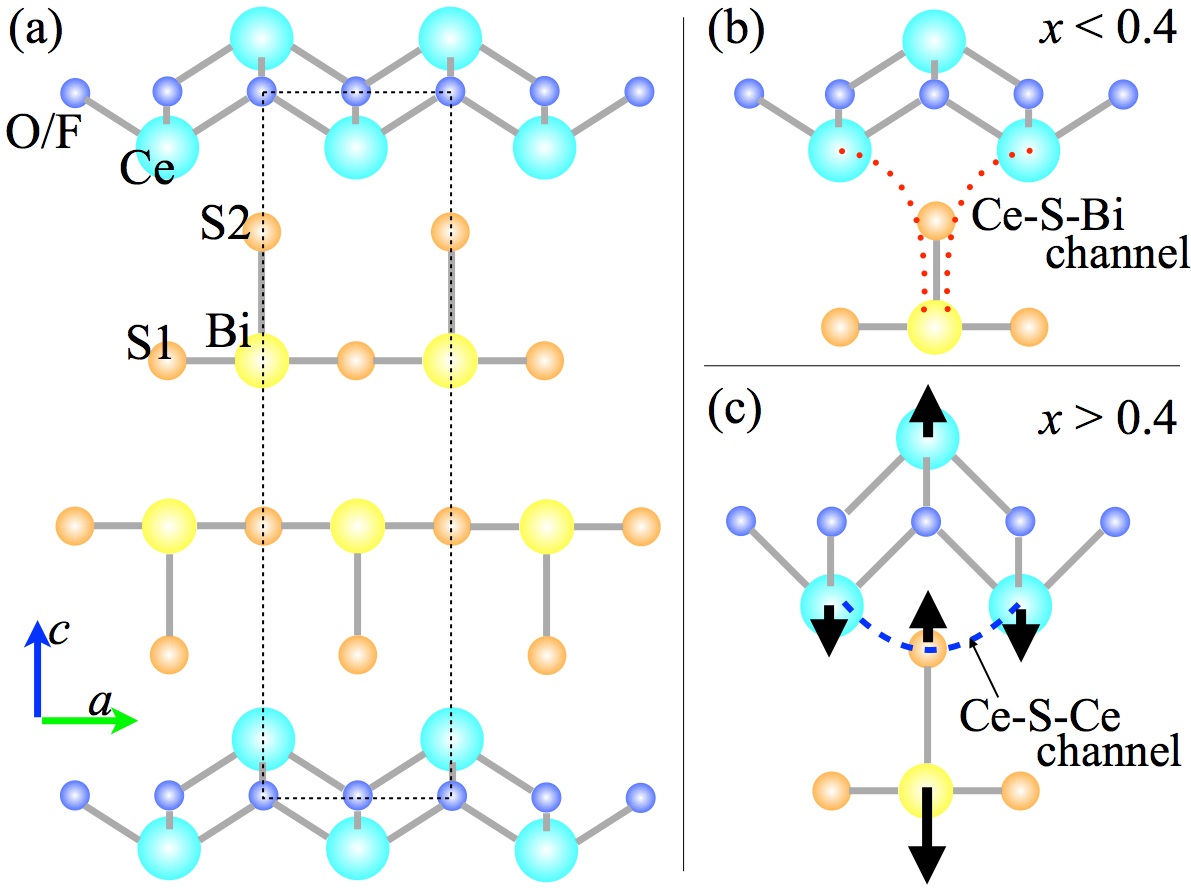}
\caption{(Color online)  
(a) The crystal structure of CeO$_{1-x}$F$_x$BiS$_2$. The dashed box represents the unit cell. (b) The local structure of CeO$_{1-x}$F$_x$BiS$_2$ for $x<0.4$ and (c) for $x>0.4$. The bond length of Ce-S2 decreases and that of Bi-S2 increases when $x>0.4$ as depicted.}
\label{Cryst}
\end{figure}

\begin{figure}
\includegraphics[width=8cm]{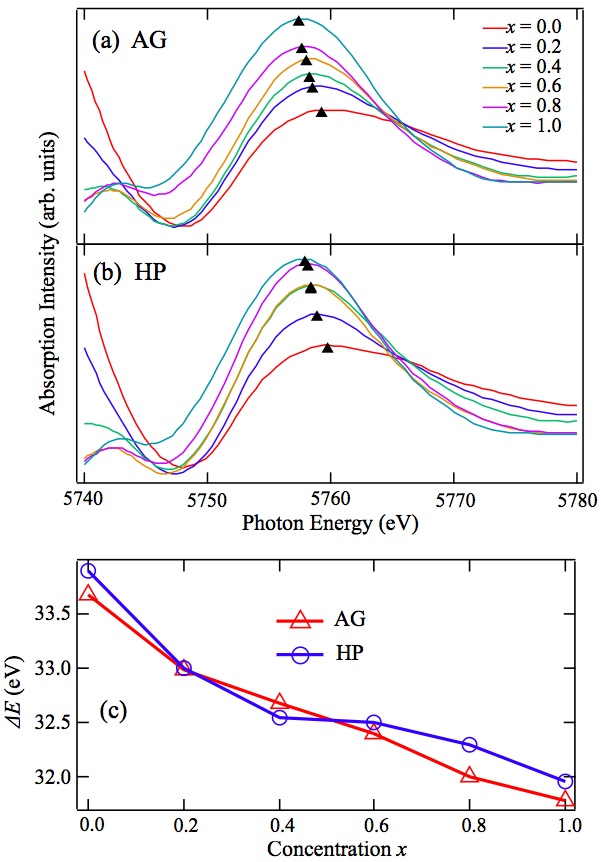}[t]
\caption{(Color online)  Evolution of the CR as a function of the F-doping 
in (a) AG samples and (b) HP samples. Peak positions are indicated by small triangles.
(c) Changes of the energy relative to the absorption white line ($\Delta E$)  as a function of the F-doping. }
\label{CR}
\end{figure}

Recalling the atomic structure, the layered structure of CeOBiS$_2$ contains  BiS$_2$ layer intercalated with CeO layer. The in-plane S atoms in the BiS$_2$ layer (S1 as in Fig. \ref{Cryst} (a)) are located at a distance of about 2.8 \AA$ $ from Bi atoms while the out-of-plane S (S2 as in Fig. \ref{Cryst}(a)) atom linking the spacer layer with the BiS$_2$ plane is located at a distance of about 2.6 \AA. By the F-substitution, the Bi-S2 distance increases ($\Delta R_{\text{max}}\sim 0.12$\AA) while Ce-S2 distance decreases ($\Delta R_{\text{max}}\sim 0.15$\AA), \cite{20,24c} leading to breaking of the Ce-S-Bi coupling channel as in Fig. \ref{Cryst} (b) to (c). Consequently, the hybridization between the Ce 4$f$ orbital and the Bi 6$p$ conduction band decreases with the F-doping and the Ce$^{3+}$/Ce$^{4+}$ valence fluctuation is suppressed. This means that the F-doping drives the system from the Ce$^{3+}$/Ce$^{4+}$ valence fluctuation regime to the Ce$^{3+}$ Kondo-like regime.\cite{22} Indeed, the Ce $4f^0$ spectral weight almost disappears at $x>0.4$, where the system shows the superconductivity and the ferromagnetism. Therefore, it seems that the mixed valence of Ce, namely, the coupling between the Ce 4$f$ and Bi 6$p$ states is not good for the superconducting state.
The Ce 4$f$ electrons are localized in the Kondo regime for $x > 0.4$ and  
that should be responsible for the ferromagnetism. The ferromagnetic Ce compounds
in the Kondo regime are rather uncommon \cite{25,26,27,28,29,30,31} and
the direct Ce-Ce exchange interaction plays an important role in many cases.
In CeO$_{1-x}$F$_x$BiS$_2$, the Ce-Ce distance is relatively longer than the ferromagnetic Ce compounds. Therefore, instead of the direct Ce-Ce exchange interaction, the Ce-S-Ce superexchange interaction as in Fig. \ref{Cryst} (c) between Ce$^{3+}$ sites is expected to be responsible for the ferromagnetism of the CeO$_{1-x}$F$_{x}$ layer. For $x > 0.4$, the ferromagnetic CeO$_{1-x}$F$_{x}$ layer is expected to be insulating just like the LaO$_{1-x}$F$_{x}$ layer, and is decoupled from the superconducting BiS$_2$ layer.

Above arguments are consistent with the structural changes observed in the same experiment through the CR peak. The CR peak is due to the Ce-Bi scattering with a contribution from the Ce-Ce. Fig. \ref{CR} shows a zoom over of the CR peak for the two series of CeO$_{1-x}$F$_x$BiS$_2$ samples, with the peak position evolving with the F-doping. The decrease of energy separation between the white line and the CR peak suggests that the Ce-Bi distance is getting elongated following the empirical rule for the CR ($\Delta E\propto1/d^2$).\cite{20a} The elongation of the Ce-Bi bond length would contribute to the reduction of the Ce-Bi coupling through Ce-S-Bi channel. Namely, the hybridization between the Ce 4$f$ orbital and the Bi 6$p$ conduction band decreases with the F-doping and the Ce$^{3+}$/Ce$^{4+}$ valence fluctuation is suppressed. This is consistent with the decrease of the $4f^0/[4f^1+4f^0]$ value by the F-doping.

%\section{Conclusion}

In conclusion, the Ce $L_3$-edge XAS results on CeO$_{1-x}$F$_x$BiS$_2$ shows that
the superconductivity and ferromagnetism are suppressed for $x < 0.4$ in the valence fluctuation regime evidenced by the $4f^1$ and $4f^0$ states.
The peak position of the CR depends on the F-doping 
and indicates the Ce-Bi bond length increases with the F-doping.
The present experimental results show that CeOBiS$_2$ has the Ce$^{3+}$/Ce$^{4+}$ valence fluctuation 
due to the Ce-Bi coupling through Ce-S-Bi channel and that the superconductivity
of the BiS$_2$ layer tends to be suppressed by the Ce-Bi coupling.
The Ce-S-Bi coupling channel is broken by the F-doping. Consequently, 
the system undergoes a crossover from the valence fluctuation regime for $x < 0.4$ 
to the Kondo-like regime for $x > 0.4$ in which superconducting BiS$_2$ layer 
and the ferromagnetic CeO$_{1-x}$F$_x$ layer are decoupled and hence coexist. The results provide important information on the role of Ce valence in the coexisting superconductivity and ferromagnetism in the CeO$_{1-x}$F$_x$BiS$_2$ system.

%\section*{Acknowledgements}

The present work is a part of the bilateral agreement between the
Sapienza University of Rome and the University of Tokyo.
The present work is partly supported by a Grant-in-Aid for 
Scientific Research from the Japan Society for Promotion of Science.

\end{document}